\documentclass[prc,onecolumn,showpacs,amsmath,amssymb, 12pt]{revtex4}

\usepackage{graphicx}
\usepackage{bm}

\begin{document}

\title{Radiation Dominated Universe for Jordan-Brans-Dicke Cosmology}

\author{M.~Ar{\i}k$^{1}$, L.~Amon Susam$^{2}$}

\address{$^{1}$Department of Physics, Bo$\breve{g}$azi\c{c}i University, Bebek, $34342$, Istanbul, Turkey\\
$^{2}$Department of Physics, Istanbul University, Istanbul, Turkey}

\begin{abstract}
Jordan-Brans-Dicke cosmology with a standard kinetic term for
the scalar field and no mass term has the same radiation dominated solution as
standard Einstein cosmology without the cosmological constant. Because of this, the
primordial nucleosynthesis (Big - Bang nucleosynthesis) result obtained for
standard cosmology remains the same for Jordan-Brans-Dicke cosmology. We show
that Jordan-Brans-Dicke cosmology with a mass term for the scalar field as well as explaining
dark energy for the present era, can also explain radiation dominated cosmology for the primordial
nucleosynthesis era.
\end{abstract}


\maketitle

\section{Introduction}

Big Bang Nucleosynthesis (BBN) is a theory based on the Standard Model which also
has some observational proof that a few minutes after the Big-Bang firstly the
stable low mass nuclei such as $^2H$, $^3He$, $^4He$, $^6Li$, $^7Li$ started to form [1-5].
This era is known as the radiation dominated nucleosynthesis period of the Universe.
So far the observational data confirm that universe is accelerating but the
reasons are not exactly known [6-8]. In cosmology it is agreed that Freidmann-Lemaitre spacetime
is used for the metric of universe [9]. General Relativity
(GR) is successful in easily explaining the accelerated expansion by introducing a
cosmological constant. This is achieved by using the Friedmann equations which
Friedmann derived from Einstein's GR by considering the possibility of an expanding universe [10-11].

As well as GR there have been various studies done by considering the Jordan-Brans-Dicke
(JBD) theory [12-15] for the acceleration of universe [16-19]. Friedmann equation for JBD scalar
tensor theory of gravitation has been derived [20] and it has been shown that JBD can make a
correction to the matter density component of the Friedmann equation.
The main parameter of JBD is $\omega$ which is a dimensionless constant and bigger
than $10^{4}$ [21]. One attractive and simple possibility is to consider a JBD cosmology
based on a scalar field
with a standard kinetic term and a mass term in the action.
Previously a work done investigated the possibility of the primordial inflation in JBD
cosmology for a closed universe and for a massive scalar field [22] which joint smoothly
with the early radiation dominated era.
In this paper we show the radiation dominated solution for JBD.

The Friedmann equation is a relation between the Hubble expansion rate $H=\frac{\dot{a}}{a}$ of
the universe and the energy density. For GR with cosmologial constant, flat, space-like sections can be expressed as

\begin{equation}
\left(\frac{H}{H_{0}}\right)^{2}= \Omega_{\bigwedge}+
\Omega_{M}\left(\frac{a_{0}}{a}\right)^{3}+\Omega_{rad}\left(\frac{a_{0}}{a}\right)^{4}
\end{equation}

where $\Omega_{\bigwedge}$ is the density parameter for dark energy, $\Omega_{M}$
is the density parameter for matter including dark matter and $\Omega_{rad}$ is density
parameter for radiation and in total they add up to unity. $\emph{a}$ is the scale size
of the universe where $\emph{H}_{0}$ and $\emph{a}_{0}$ denotes to today values. The radiation
contribution is taken negligible for this work.

The standard JBD action in the canonical form is given by

\begin{equation}
S = \int d^{4}x \sqrt{g}\:\left[-\frac{1}{8\omega}\phi^{2}R
+\frac{1}{2}g^{\mu\nu}\partial_{\mu}\phi\partial_{\nu}\phi
-\frac{1}{2}m^{2}\phi^{2}+ L_{M}\right].
\end{equation}

The metric signature is (+ - - -). $g$ is minus the determinant of the metric
and $d^{4}x\sqrt{g}$ is the four-dimensional volume form, $\phi$ is the scalar field,
$R$ is the Ricci scalar, $g^{\mu\nu}$ is the metric tensor and $L_{M}$ is the matter
Lagrangian density except the scalar field $\phi$. Note that the standard mass term for the scalar field can
be combined with the curvature term so that this is sometimes called the cosmological term for
JBD theory [23-26].The field equations derived from the variation of the action (2) with
respect to Robertson - Walker metric are

\begin{equation}
\frac{3}{4\omega}\phi^{2}\left(\frac{\dot{a}^{2}}{a^{2}}
+ \frac{k}{a^{2}}\right)- \frac{1}{2}\dot{\phi}^{2} - \frac{1}{2}m^{2}\phi^{2}
+\frac{3}{2\omega}\frac{\dot{a}}{a}\dot{\phi}\phi = \rho_{M}
\end{equation}

\begin{equation}
-\frac{1}{4\omega}\phi^{2}\left(2\frac{\ddot{a}}{a}+\frac{\dot{a}^{2}}{a^{2}}
+ \frac{k}{a^{2}}\right)-\frac{1}{\omega}\frac{\dot{a}}{a} \dot{\phi}\phi -
\frac{1}{2\omega}\ddot{\phi}\phi\ \left(\frac{1}{2} +\frac{1}{2\omega}\right)\dot{\phi}^{2}
+ \frac{1}{2}m^{2}\phi^{2} = p_{M}
\end{equation}

\begin{equation}
\ddot{\phi} + 3\frac{\dot{a}}{a}\dot{\phi} + \left[m^{2}
- \frac{3}{2\omega}\left( \frac{\ddot{a}}{a}
+ \frac{\dot{a}^{2}}{a^{2}} +\frac{k}{a^{2}}\right)\right]\phi = 0
\end{equation}

where \textit{k} is the curvature parameter which \textit{k} = -1, 0, 1 correspond
to open, flat, closed universes respectively, \emph{a}(\textit{t}) is the scale
factor of the universe, $\dot{a}(\textit{t}) $ denotes the derivative of that factor and $M$
denotes everything except the scalar field. For
k=0, m=0, $\emph{a}\sim$t$^{\alpha}$, $\phi\sim$t$^{\beta}$, $\rho\sim$a$^{-4}$
and $p=\rho/3$ in (3-5) it gives

\begin{equation}
a\sim t^{1/2} , \phi=constant.
\end{equation}

Since $m$ is different than zero for dark energy, to generalize this solution
a more general treatment is needed. Instead of working with the field equations
(3-5) which are given in terms of
$\phi(\textit{t}), \emph{a}(\textit{t})$ and their derivatives with respect to
cosmological time \textit{t}, the fractional rate of change of $\phi$
and $\emph{a}$ is taken as

\begin{equation}
F(\emph{a}) =\frac{\dot{\phi}}{\phi}
\end{equation}

and the Hubble parameter is taken as

\begin{equation}
H(\emph{a})= \frac{\dot{a}}{a},
\end{equation}

and rewritten the left hand side of the field equations in terms of $H(\emph{a})$, $F(\emph{a})$
and their derivatives with respect to $\emph{a}$

\begin{equation}
H^{2} - \frac{2\omega}{3} F^{2} + 2 H F + \frac{k}{{\emph{a}}^{2}}
-\frac{2\omega}{3}m^{2}= \left(\frac{4\omega}{3}\right) \frac{\rho}{\phi^{2}}
\end{equation}

\begin{equation}
H^{2} + \left(\frac{2\omega}{3} + \frac{4}{3}\right) F^{2}
+ \frac{4}{3} H F +\frac{2\emph{a}}{3} \left(H H^{'} + H F^{'}\right)+
\frac{k}{3\emph{a}^{2}}- \frac{2\omega}{3}m^{2}
= \left(-\frac{4\omega}{3}\right)\frac{p}{\phi^{2}}
\end{equation}

\begin{equation}
H^{2} - \frac{\omega}{3}F^{2} - \omega H F + \emph{a}\left( \frac{H H^{'}}{2}
-\frac{\omega}{3} H F^{'} \right) + \frac{k}{2\emph{a}^{2}} -\frac{\omega}{3}m^{2}=0.
\end{equation}

From these three equations it is shown that the continuity equation for
the matter - energy excluding the JBD scalar field is as follows

\begin{equation}
\dot{\rho}+3\left(\frac{\dot{\emph{a}}}{\emph{a}}\right) \left(p +\rho\right) = 0.
\end{equation}

The continuity equation is used in place of (10) because (9,11 and 12) is
used to derive (10) provided $\rho\neq0$. After some arrangements equation (9)
  is received purely in terms of $H(\emph{a})$, $F(\emph{a})$, $\rho(\emph{a})$ and their
derivatives with respect to \emph{a}.

\begin{eqnarray}
H^{'}\left(H^{2}+HF\right)+F^{'}\left(H^{2}-\frac{2\omega}{3}HF\right)=
\frac{H^{3}}{2}\left(\frac{\rho^{'}}{\rho}\right)+
\frac{2\omega}{3\emph{a}}F^{3}+ H^{2}F\left[\left(\frac{\rho^{'}}{\rho}\right)
-\frac{1}{\emph{a}}\right]\nonumber
\end{eqnarray}
\begin{eqnarray}
+F^{2}H \left[-\frac{2}{\emph{a}}
-\frac{\omega}{3}\left(\frac{\rho^{'}}{\rho} \right) \right] +\frac{k}{\emph{a}^{2}} \left[H\left(\left(\frac{\rho^{'}}{2\rho}\right)
+\frac{1}{\emph{a}}\right)-\frac{F}{\emph{a}}\right]
- \omega m^{2}\left[H\left(\frac{\rho^{'}}{3\rho}\right )
-\frac{2F}{3\emph{a}}\right]
\end{eqnarray}

After rewriting (11) in the following form

\begin{equation}
3\emph{a} H H^{'} - 2\omega H \emph{a} F^{'} = -6H^{2}
+ 2\omega F^{2}+ 6\omega HF
- \frac{3k}{\emph{a}^{2}} + 2\omega m^{2}
\end{equation}

(13,14) are solved for $H^{'}$, $F^{'}$

\begin{eqnarray}
H^{'} =\frac{\left[\omega\emph{a}\left(\frac{\rho^{'}}{\rho}\right)
-6\right]}{\left(2\omega+3\right)\emph{a}H}H^{2}
-\frac{\left[4\omega^{2}+2\omega
+2\emph{a}\omega^{2}\left(\frac{\rho^{'}}{3\rho}\right)\right]}
{\left(2\omega+3\right)\emph{a}H}F^{2}+
\frac{\left[8\omega+2\emph{a}\omega\left(\frac{\rho^{'}}{\rho}\right)\right]}{\left(2\omega+3\right)\emph{a}H}HF\nonumber
\end{eqnarray}
\begin{eqnarray}
-\frac{\left[2\omega^{2}\emph{a}\left(\frac{\rho^{'}}{3\rho}\right)-2\omega\right]}{\left(2\omega+3\right)\emph{a}H}m^{2}+k \frac{\left[2\omega +\omega\emph{a}\left(\frac{\rho^{'}}{2}\right)-3\right]}{\left(2\omega+3\right)\emph{a}^{3}H}
\end{eqnarray}

\begin{eqnarray}
F^{'}=\frac{\left[3\emph{a}\left(\frac{\rho^{'}}{2\rho}\right)+6\right]}{\left(2\omega+3\right)\emph{a}H}H^{2}
-\frac{\left[8\omega+\emph{a}\omega\left(\frac{\rho^{'}}{\rho}\right)+6\right]}{\left(2\omega+3\right)\emph{a}H}F^{2}-
\frac{\left[6\omega-3\emph{a}\left(\frac{\rho^{'}}{\rho}\right)-3\right]}{\left(2\omega+3\right)\emph{a}H}HF\nonumber
\end{eqnarray}
\begin{eqnarray}
-
\frac{\left[\omega\emph{a}\left(\frac{\rho^{'}}{\rho}\right)+2\omega\right]}{\left(2\omega+3\right)
\emph{a}H}m^{2}+k\frac{\left[6+3\emph{a}\left(\frac{\rho^{'}}{2\rho}\right)\right]}{\left(2\omega+3\right)\emph{a}^{3}H}.
\end{eqnarray}

Using (12), it is seen that the energy density $\rho$ evolves with \emph{a} in the
same manner as in standard Einstein cosmology when the universe is solely
governed by radiation,

\begin{equation}
\rho=\frac{C}{\emph{a}^{4}}
\end{equation}

where \textit{C} is an integration constant.

\section{A Perturbative Method to Find The Solution For $m\neq0$}

To discuss the solutions of (9-11) firstly the vacuum solution is considered where
$\rho$ and $p$ are taken to be equal to zero and $k=0$ set for flat, space-like sections.

\begin{equation}
H^{2} - \frac{2\omega}{3} F^{2} + 2 H F-\frac{2\omega}{3}m^{2}=0
\end{equation}

\begin{equation}
H^{2} + \left(\frac{2\omega}{3} + \frac{4}{3}\right) F^{2}
+ \frac{4}{3} H F-\frac{2\omega}{3}m^{2}
=0
\end{equation}

\begin{equation}
H^{2} - \frac{\omega}{3}F^{2} - \omega H F -\frac{\omega}{3}m^{2}=0.
\end{equation}

The terms with $m$ are carried to the right hand side of the equations and (20) is multiplied
by a factor of $2$. A solution such as $H=H_{0}$ and $F=F_{0}$ is considered where $H$ and $F$ are constant.

\begin{equation}
H_{0}^{2} - \frac{2\omega}{3} F_{0}^{2} + 2 H_{0} F_{0}
=\frac{2\omega}{3}m^{2}
\end{equation}

\begin{equation}
H_{0}^{2} + \left(\frac{2\omega}{3} + \frac{4}{3}\right) F_{0}^{2}
+ \frac{4}{3} H_{0} F_{0}=\frac{2\omega}{3}m^{2}
\end{equation}

\begin{equation}
H_{0}^{2} - \frac{\omega}{3}F_{0}^{2} - \omega H_{0} F_{0}
=\frac{\omega}{3}m^{2}.
\end{equation}

Since the left hand side equations are homogeneous in $H_{0}$ and $F_{0}$, $H_{0}=cF_{0}$ is
taken and it is found that

\begin{eqnarray}
c=2\omega+2,\nonumber
\end{eqnarray}
\begin{eqnarray}
H_{0}= \sqrt{\frac{(4\omega^{2}+8\omega+4)\omega m^{2}}{6\omega^{2}+17\omega+12}},
\end{eqnarray}
\begin{eqnarray}
F_{0}= \frac{H_{0}}{2\omega+2}.\nonumber
\end{eqnarray}

To consider how radiation modifies this solution, $\rho=3p$ is taken.
Two equations are obtained from the main three equations (9-11).

\begin{equation}
\frac{4H^{2}}{3}+\left(\frac{4\omega}{9}+
\frac{4}{3}\right)F^{2}+2HF+\frac{2\emph{a}}{3}\left(HH'+HF'\right)
-\frac{8\omega}{9}m^{2}=0
\end{equation}

\begin{equation}
H^{2}-\frac{\omega}{3}F^{2}-\omega HF+\emph{a}\left(\frac{HH'}{2}-\frac{\omega}{3}HF'\right)
-\frac{\omega}{3}m^{2}=0
\end{equation}

To investigate a perturbative solution $H=H_{0}+H_{1}\emph{a}^{n}+...$ and $F=F_{0}+F_{1}\emph{a}^{n}+...$
are taken where dots denote terms of $\emph{a}^{2n}$ and higher, and  $n$ is determined from the above equations.
The equations below are obtained.

\begin{eqnarray}
\emph{a}^{2n}\left[H_{1}^{2}\left(1+\frac{n}{2}\right)-F_{1}^{2}\frac{\omega}{3}-\omega H_{1}F_{1}
\left(1+\frac{n}{3}\right)\right]\nonumber
\end{eqnarray}
\begin{eqnarray}
+\emph{a}^{n}\left[H_{0}H_{1}\left(2+\frac{n}{2}\right)-\frac{2\omega}{3}F_{0}F_{1}-\omega H_{0}F_{1}
\left(1+\frac{n}{3}\right)-\omega H_{1}F_{0}\right]+\nonumber
\end{eqnarray}
\begin{eqnarray}
\left[H_{0}^{2}-\frac{\omega}{3}F_{0}^{2}-\omega H_{0}F_{0}-\frac{\omega}{3}m^{2}\right]=0
\end{eqnarray}

\begin{eqnarray} \emph{a}^{2n}\left[\frac{H_{1}^{2}}{4}\left(2+n\right)+\frac{F_{1}^{2}}{2}\left(\frac{\omega}{3}+1\right)
+\frac{H_{1}F_{1}}{4}\left(3+n\right)\right]\nonumber
\end{eqnarray}
\begin{eqnarray}
+\emph{a}^{n}\left[F_{0}F_{1}\left(\frac{\omega}{3}+1\right)+H_{0}H_{1}\left(1+\frac{n}{4}\right)
+\frac{H_{0}F_{1}}{4}\left(3+n\right)+H_{1}F_{0}\frac{3}{4} \right]+\nonumber
\end{eqnarray}
\begin{eqnarray}
\left[\frac{H_{0}^{2}}{2}+\frac{F_{0}^{2}}{2}\left(\frac{\omega}{3}+1\right)+H_{0}F_{0}\frac{3}{4}
-\frac{\omega}{3}m^{2}\right]=0
\end{eqnarray}

Setting the coefficients of $\emph{a}^{n}$ equal to zero we obtain

\begin{eqnarray}
H_{0}H_{1}\left(2+\frac{n}{2}\right)-\frac{2\omega}{3}F_{0}F_{1}-\omega H_{0}F_{1}\left(1+\frac{n}{3}\right)
-\omega H_{1}F_{0}=0
\end{eqnarray}

\begin{eqnarray}
F_{0}F_{1}\left(\frac{\omega}{3}+1\right)+H_{0}H_{1}\left(1+\frac{n}{4}\right)+
\frac{H_{0}F_{1}}{4}\left(3+n\right)+H_{1}F_{0}\frac{3}{4}=0
\end{eqnarray}

$H_{0}=cF_{0}$ is inserted, then


\begin{eqnarray}
H_{1} \left(2 - \frac{\omega}{c} + \frac{n}{2} \right) - F_{1} \left( \frac{2 \omega}{3 c} + \omega + \frac{\omega n}{c} \right)=0
\end{eqnarray}

\begin{eqnarray}
H_{1} \left( 1 + \frac{3}{4 c} +\frac{n}{4} \right) + F_{1} \left( \frac{\omega}{3 c} + \frac{1}{c} + \frac{3}{4} + \frac{n}{4} \right)=0
\end{eqnarray}

are obtained. For a nontrivial solution the determinant of the coefficients must vanish and two solutions are obtained.

\begin{eqnarray}
H_{1}=- \frac{2 F_{1}}{3}
\end{eqnarray}

and

\begin{eqnarray}
H_{1}= \frac{2 \omega F_{1}}{3}.
\end{eqnarray}

For the solution
$H_{1}=- \frac{2 F_{1}}{3}$; $n= -\frac{3 \omega +4}{\omega + 1}$ is obtained from (20) and (21)
whereas for $H_{1}=\frac{2\omega F_{1}}{3}$, $n= -\frac{4\omega +5}{\omega + 1}$ is obtained.
Since standard radiation dominated cosmology with cosmological constant is
$H = H_{0} + H_{1} \emph{a}^{-4}$ only the solution $F_{1}=\frac{3 H_{1}}{ 2\omega}$ ,
$n=-\frac{4\omega +5}{\omega + 1} \approx -4$ is considered.
\\

Then we move one step further and we search for higher order terms. The equations,
$ H = H_{0} + H_{1} \emph{a}^{n} + H_{2} \emph{a}^{2n}+...$ and
$ F = F_{0} + F_{1} \emph{a}^{n} + F_{2} \emph{a}^{2n}+...$ are used in (25) and (26).
Two equations which contains $\emph{a}^{2n}$, $\emph{a}^{3n}$, $\emph{a}^{4n}$ are
obtained and only the lowest order $\emph{a}^{2n}$ are taken from these equations.
\\
\begin{eqnarray}
\emph{a}^{2n}[\frac{4\omega^{2}}{9}F_{1}^{2} \left(\frac{4}{3}+\frac{2n}{3}\right)+
F_{1}^{2}\left(\frac{4}{3}+\frac{4\omega}{9} \right) + (2\omega+2) H_{2}F_{0} \left(\frac{4n}{3}+\frac{8}{3} \right) \nonumber
\end{eqnarray}
\begin{eqnarray}
+ F_{2}F_{0} \left(\frac{8}{3}+\frac{8\omega}{9} \right)+(2\omega+2) F_{2}F_{0}\left(\frac{4n}{3}+2 \right)
+ \frac{2\omega}{3} F_{1}^{2}\left(\frac{2n}{3}+2 \right) \nonumber
\end{eqnarray}
\begin{eqnarray}
+ 2H_{2}F_{0}]=0
\end{eqnarray}

\begin{eqnarray}
\emph{a}^{2n} [\frac{4\omega^{2}}{9}F_{1}^{2}\left(1+\frac{n}{2}\right) -\frac{\omega}{3} F_{1}^{2} +(2\omega+2) H_{2}F_{0}(2+n)
- \frac{2\omega}{3}F_{2}F_{0} \nonumber
\end{eqnarray}
\begin{eqnarray}
- (2\omega+2) F_{2}F_{0} \left(\omega+\frac{2\omega n}{3}\right)  - \frac{2\omega}{3} F_{1}^{2}\left(\omega+\frac{\omega n}{3}\right)
- \omega H_{2}F_{0}]=0
\end{eqnarray}

(35) and (36) are separated in parentheses of $H_{0}$ and $F_{0}$.

\begin{eqnarray}
\left(\frac{4}{3}H_{2} + \frac{2 n}{3} H_{2} +  F_{2} + \frac{2 n}{3} F_{2} \right) H_{0} + \left(\frac{4\omega}{9}F_{2} + \frac{4}{3}F_{2} +  H_{2} \right) F_{0} = \nonumber \\ \left( -\frac{2}{3} H_{1}^{2} - \frac{2\omega}{9} F_{1}^{2} -\frac{2}{3} F_{1}^{2} -  H_{1} F_{1} -\frac{ n}{3} H_{1}^{2} - \frac{ n}{3} H_{1} F_{1}\right)
\end{eqnarray}

\begin{eqnarray}
\left(2 H_{2} + n H_{2} - \omega F_{2} - \frac{2\omega n}{3} F_{2}  \right) H_{0} + \left(- \frac{2\omega}{3}F_{2} - \omega H_{2}  \right) F_{0} = \nonumber \\ \left( - H_{1}^{2} + \frac{\omega}{3} F_{1}^{2} + \omega H_{1} F_{1} -\frac{ n}{2} H_{1}^{2} + \frac{\omega n}{3} H_{1} F_{1}\right)
\end{eqnarray}

The results for $H_{2}$ and $F_{2}$ are obtained as below where terms which are lower order in $\omega$ are neglected.

\begin{eqnarray}
H_{2} \approx - \frac{ H_{1}^{2}}{ H_{0}} \left [ \frac{1}{2} + \frac{6}{5 \omega} \right]
\end{eqnarray}

and

\begin{eqnarray}
F_{2} \approx - \frac{F_{1}^{2}}{H_{0}} \frac{\omega}{3}=-\frac{3H_{1}^{2}}{4\omega H_{0}}.
\end{eqnarray}

To obtain the Friedmann equation for radiation dominated era the result for $H$ is written as

\begin{eqnarray}
H = H_{0} + H_{1} \emph{a}^{-4} - \frac{H_{1}^{2}}{H_{0}} \left [ \frac{1}{2} + \frac{6}{5 \omega} \right]  \emph{a}^{-8}+...
\end{eqnarray}

which gives

\begin{eqnarray}
H^{2} = H_{0}^{2} + 2 H_{0} H_{1} \emph{a}^{-4}  - \frac{12 H_{1}^{2} }{5 \omega} \emph{a}^{-8}+...
\end{eqnarray}

For the radiation dominated era we thus obtain a correction to the standard result obtained when Einstein-Hilbert action with cosmological constant is used.
This is important since BBN depends critically on how the scale parameter $(\emph{a})$ changes as a function of time during the radiation dominated era.
The conditions for the constant term and the $\emph{a}^{-8}$ term to be negligible give

\begin{eqnarray}
\frac{6 H_{1}}{5 \omega H_{0}} \ll  \emph{a}^{4} \ll \frac{2 H_{1}}{ H_{0}}
\end{eqnarray}

which is satisfied since $\omega\gg 10^{4} $.
Note that $H_{0}$ is given by (24) whereas  $H_{1}$ is a free parameter for our solution.
Furthermore our method gives a way to calculate the corrections
to the standard model which we have shown to be negligible.

\section{Conclusion}

JBD theory with $\textit{m}=0$ admits the
same radiation dominated solution as classical Einstein solutions.
It is important to extend this solution to the $\textit{m}\neq0$ case since
this case corresponds to dark energy. By using a power series approach to
obtain the Friedmann equation we have shown that there are two corrections
to the standard $\emph{a}^{-4}$ term. As expected, one of them is the constant
term which corresponds to dark energy. The other is the $\emph{a}^{-8}$ term which
follows from the power series expansion. We have shown that both terms may be chosen
to be small during the radiation dominated era. In fact from equation (43) it can be seen that
if the scale size $\emph{a}$ changes by a factor of $f$ during the radiation dominated
era then the Brans-Dicke parameter $\omega$ must be greater than $f^{4}$.
Our results are important because they say that JBD theory can explain
dark energy without contradicting the successful BBN predictions of the standard cosmological model.


\begin{thebibliography}{00}
\bibitem{1} C. Grupen, \textit{Astroparticle Physics}, Springer-Verlag New York, LLC (2008).
\bibitem{2}  R.V. Wagoner, W. A. Fowler and F. Hoyle  \textit{On the synthesis of the elements at very high temperatures, Astrophys. J.} \textbf{148} (1967) 3.
\bibitem{3}  D. N. Schramm and M. S. Turner,\textit{Big-bang nucleosynthesis enters the precision era, Rev. Mod. Phys.}  \textbf{70} (1998) 303.
\bibitem{4}  R.A. Malaney and G.J. Mathews,\textit{Probing the early universe - a review of primordial nucleosynthesis beyond the standard big-bang
 ,Phys. Rep.} \textbf{229} (1993) 145.
\bibitem{5}  C. Amsler et al., \textit{Review of particle physics, Phys. Lett. B} \textbf{667} (2008) 1.
\bibitem{6}  A. G. Reiss et al., \textit{Observational evidence from supernovae for an accelerating universe and a cosmological constant, Astron. J.} \textbf{116} (1998) 1009.
\bibitem{7}  J. E. Gunn and B. M. Tinsley, \textit{An accelerating universe, Nature} \textbf{257} (1975) 454.
\bibitem{8}  S. Perlmutter et al.,\textit{Measurements of omega and lambda from 42 high-redshift supernovae, Astrophys. J.} \textbf{517} (1999) 565.
\bibitem{9}  J.P. Uzan, \textit{The fundamental constants and their variation: observational and theoretical status, Rev. Mod. Phys} \textbf{75} (2003) 403.
\bibitem{10} S. M. Carroll and M. Kiplinghat,\textit{Testing the Friedmann equation: The expansion of the universe during big-bang nucleosynthesis, Phys. Rev. D} \textbf{65} (2002) 063507.
\bibitem{11} J. M. Bardeen, P. J. Steinhardt and M. S. Turner, \textit{Spontaneous creation of almost scale-free density perturbations in an inflationary universe, Phys. Rev. D} \textbf{28} (1983) 679.
\bibitem{12} C. Brans and R. H. Dicke, \textit{Mach's principle and a relativistic theory of gravitation, Phys. Rev.} \textbf{124} (1961) 925.
\bibitem{13} R. H. Dicke, \textit{Mach's principle and invariance under transformation of units, Phys. Rev.} \textbf{125} (1962) 2163.
\bibitem{14} P. Jordan, \textit{Zur empirischen Kosmologie, Naturwiss} \textbf{26} (1938) 417.
\bibitem{15} P. Jordan, \textit{The present state of Dirac's cosmological hypothesis, Z. Phys.} \textbf{157} (1959) 112.
\bibitem{16} B. Boisseau, G. Esposito-Farese, D. Polarski and A. A. Starobinsky, \textit{Reconstruction of a scalar-tensor theory of gravity in an accelerating universe, Phys. Rev. Lett.} \textbf{85} (2000) 2236.
\bibitem{17} V. Acquaviva and L. Verde, \textit{Observational signatures of Jordan�Brans�Dicke theories of gravity, J. Cosmol. Astropart. P.} \textbf{12} (2007) 1.
\bibitem{18} R.M. Avagian, G.H. Harutyunyan and W. Papoyan, \textit{Cosmological scalar in the Jordan-Brans-Dicke theory. I, Astrophysics} \textbf{48} (2005) 3.
\bibitem{19} V. Pettorino, C. Baccigalupi and G. Mangano, \textit{Extended quintessence with an exponential coupling, J. Cosmol. Astropart. P.} \textbf{1} (2005) 14.
\bibitem{20} M. Ar{\i}k, M. C. \c{C}al{\i}k and M. B. Sheftel, \textit{Friedmann equation for Brans-Dicke cosmology., Int. J. Mod. Phys. D}
\textbf{17} (2008) 225.
\bibitem{21} C. M. Will, \textit{Theory and experiment in gravitational physics}, Cambridge University Press, Cambridge UK (1993).
\bibitem{22} M. Ar{\i}k and M. C. \c{C}al{\i}k, \textit{Primordial and late-time inflation in Brans�Dicke cosmology, J. Cosmol. Astropart. P.} \textbf{01}, (2005) 013.
\bibitem{23} O. Delice, \textit{Cylindrically symmetric, static strings with a cosmological constant in Brans-Dicke theory, Phys. Rev. D} \textbf{74} (2006) 1240001.
\bibitem{24} B. Linet, \textit{The static, cylindrically symmetric strings in general relativity with cosmological constant, J. Math. Phys.} \textbf{27} (1986) 1817.
\bibitem{25} Q. Tian, \textit{Cosmic strings with cosmological constant, Phys. Rev. D} \textbf{33} (1986) 3549.
\bibitem{26} D. Lorenz - Petzold, \textit{Exact Brans-Dicke Bianchi type-I solutions with a cosmological constant, Phys. Rev. D} \textbf{29} (1984) 2399.
\end{thebibliography}
\end{document}